\documentclass[trackchanges]{aastex7}
\usepackage{makecell}

\begin{document}

\title{A neural network model for quickly solving multiple-band light curves of contact binaries}

\author[0000-0003-3590-335X]{Kai Li}
\affiliation{Shandong Key Laboratory of Space Environment and Exploration Technology, School of Space Science and Technology, Institute of Space Sciences, Shandong University, Weihai, Shandong, 264209, China}
\email[show]{kaili@sdu.edu.cn}

\author[0009-0005-0485-418X]{Li-Heng Wang}
\affiliation{Shandong Key Laboratory of Space Environment and Exploration Technology, School of Space Science and Technology, School of Space Science and Technology, Institute of Space Sciences, Shandong University, Weihai, Shandong, 264209, China}
\email{kaili@sdu.edu.cn}

\author[0009-0009-6364-0391]{Xiang Gao}
\affiliation{Shandong Key Laboratory of Space Environment and Exploration Technology, School of Space Science and Technology, School of Space Science and Technology, Institute of Space Sciences, Shandong University, Weihai, Shandong, 264209, China}
\email{kaili@sdu.edu.cn}

\begin{abstract}
The advent of large-scale photometric surveys has led to the discovery of over a million contact binary systems. Conventional light curve analysis methods are no longer adequate for handling such massive datasets. To address this challenge, we developed a neural network-based model capable of rapid analysis of multiple-band light curves of contact binaries. Our model can determine the fundamental physical parameters, including temperature and mass ratios, orbital inclination, potential, fillout factor, primary and secondary luminosities and radii, third light contribution, and spot parameters. Notably, unlike previous works, our model can simultaneously process multiple-band light curves and the four parameters of a starspot. The model's reliability was verified through analysis of the synthetic light curves generated by PHOEBE and the light curves of eight targets from \cite{2024ApJ...976..223W}'s work. The discrepancy distribution between the physical parameters determined by our model and true values for the synthetic light curves shows very good agreement. In addition,
the physical parameters determined by our model and the corresponding light curve fits show remarkable consistency with \cite{2024ApJ...976..223W}'s results. By applying our model to OGLE contact binaries, physical parameters of 3,541 systems were obtained. We have packaged our model into an executable (CBLA.exe) file and archived it in the China-VO repository (\url{https://doi.org/10.12149/101626}). The software supports light curve analysis of 19 standard filters and
allows for processing the data, whether from large-scale sky surveys or individual telescope observations.

\end{abstract}


\section{Introduction} 
Contact binaries are the most numerous type of close binary star; approximately 1 in every 500 main-sequence stars in our Galaxy is a contact binary \citep{2007MNRAS.382..393R}. Their two components often differ significantly in mass, while their temperatures are very similar \citep{1941ApJ....93..133K}. They typically consist of two late-type stars who share a common envelope with orbital periods typically under one day and serve as critical astrophysical laboratories with multifaceted importance. Combined light and radial velocity curve analyses yield fundamental parameters of stars, making them vital benchmarks for calibrating stellar models \citep{1981ApJ...245..650M,2003A&A...412..465K}. W UMa-type contact binaries display a well-defined period-luminosity relationship, making them valuable standard candles for distance measurements from the Local Group and beyond (e.g., \citealt{1994PASP..106..462R,2016ApJ...832..138C,2018ApJ...859..140C}). These binaries also serve as laboratories for testing binary evolution theory: their continuous mass transfer through the inner Lagrangian point and angular momentum loss provide empirical validation for key processes, including common-envelope phase \citep{2011A&A...528A.114T,2013A&ARv..21...59I}.

Since the 1940s, significant progress has been made in research on contact binary stars \citep{1941ApJ....93..133K}. Nevertheless, many issues remain to be addressed, including their formation, evolution, and final fate \citep{1994ASPC...56..228B,2002ApJ...575..461E,2005ApJ...629.1055Y,2024AJ....168..272P}, the O'Connell effect \citep{1951PRCO....2...85O}, the mass ratio limit \citep{1995ApJ...444L..41R,2006MNRAS.369.2001L,2024A&A...692L...4L},  and the short period cut-off \citep{1992AJ....103..960R,2012MNRAS.421.2769J,2019MNRAS.485.4588L}. The key to addressing these issues lies in obtaining accurate physical parameters for a large sample of contact binary stars.

In recent years, several astronomical surveys have significantly expanded our knowledge of contact binary stars. Notably, the Catalina Sky Survey (CSS; \citealt{2024ApJS..273...31W}), the All Sky Automated Survey for SuperNovae (ASAS-SN, \citealt{2018MNRAS.477.3145J}), the Zwicky Transient Facility (ZTF, \citealt{2020ApJS..249...18C}), the Transiting Exoplanet Survey Satellite (TESS, \citealt{2022ApJS..258...16P}), and the Gaia mission \citep{2023A&A...674A..16M} have identified over one million contact binary stars. Traditional light curve (LC) analysis methods, including Wilson-Devinney (W-D) code \citep{1971ApJ...166..605W, 1979ApJ...234.1054W, 1990ApJ...356..613W} and PHysics Of Eclipsing BinariEs (PHOEBE; \citealt{2005ApJ...628..426P,2016ApJS..227...29P}) software, can hardly accomplish the analysis of such a large number of contact binaries. Machine learning methods offer the possibility of analyzing massive LC samples. \cite{2022AJ....164..200D} developed a machine learning model that incorporates the Markov Chain Monte Carlo (MCMC) algorithm \cite{2019JOSS....4.1864F} to rapidly derive the physical parameters of contact binaries. This model was trained using PHOEBE generated LCs and subsequently applied to LCs from three photometric surveys, TESS \citep{2023MNRAS.525.4596D}, ASAS-SN \citep{2024ApJS..271...32L}, and CSS \citep{2024ApJS..273...31W}. Very recently, \cite{2025ApJS..277...51L} improved Ding et al.'s model by incorporating spot parameters to derive more accurate physical parameters. Although some progress has been made in analyzing LCs with machine learning methods, current models remain unable to simultaneously analyze multiple-band observations and optimize all four spot parameters. Therefore, this paper presents a neural network (NN) model, trained on PHOEBE-generated LCs, which is capable of simultaneously analyzing multiple-band LCs and optimizing all four spot parameters.

\section{Establishment of Neural Network Model} 
We used a NN to establish a mapping relationship between the physical parameters and the LCs in phased normalized flux. The LCs for training are generated using PHOEBE. The physical parameters and their ranges used to generate LCs are as follows: the effective temperature of the primary component ($T_1$, from 4000K to 10000K), the temperature ratio ($T_2/T_1$, from 0.7 to 1.2), the mass ratio (q, from 0 to 1), the orbital inclination ($i$, from $30^\circ$ to $90^\circ$), the contact degree (f, from 0 to 1), the third light ($l_3$, from 0 to 0.6), and the spot parameters, latitude ($\theta$, from $0^\circ$ to $180^\circ$), longitude ($\lambda$, from $0^\circ$ to $360^\circ$), angular radius($r_s$, from $0^\circ$ to $60^\circ$), and relative temperature ($T_s$, from 0.6 to 1.4). The atmospheric model proposed by \cite{2004A&A...419..725C} and the V-band were applied in our model. If the temperature is less than 7200 K, the gravity darkening and bolometric albedo coefficients were set as $g=0.32$ \citep{1967ZA.....65...89L} and $A=0.5$ \citep{1969AcA....19..245R}, setting both parameters to 1 when the temperature exceeds 7200 K.
Then, we constructed the full parameter space and performed uniform sampling to generate normalized flux LCs using PHOEBE. 500,000 parameter sets were used. Based on each parameter set, we calculated the luminosity ($L_1$, $L_2$), radius ($R_1$, $R_2$), and potential ($\Omega_1=\Omega_2$) of each component. Finally, a total of 437,202 parameter sets and corresponding LCs were generated after excluding those that were inconsistent with the contact binary model. During training, the dataset was divided into training and validation sets in an 8:2 ratio. After completion of model training, 5000 parameter sets were randomly sampled to generate LCs using PHOEBE, which established the test set.

\begin{figure*}
\centering
\includegraphics[width=0.6\textwidth]{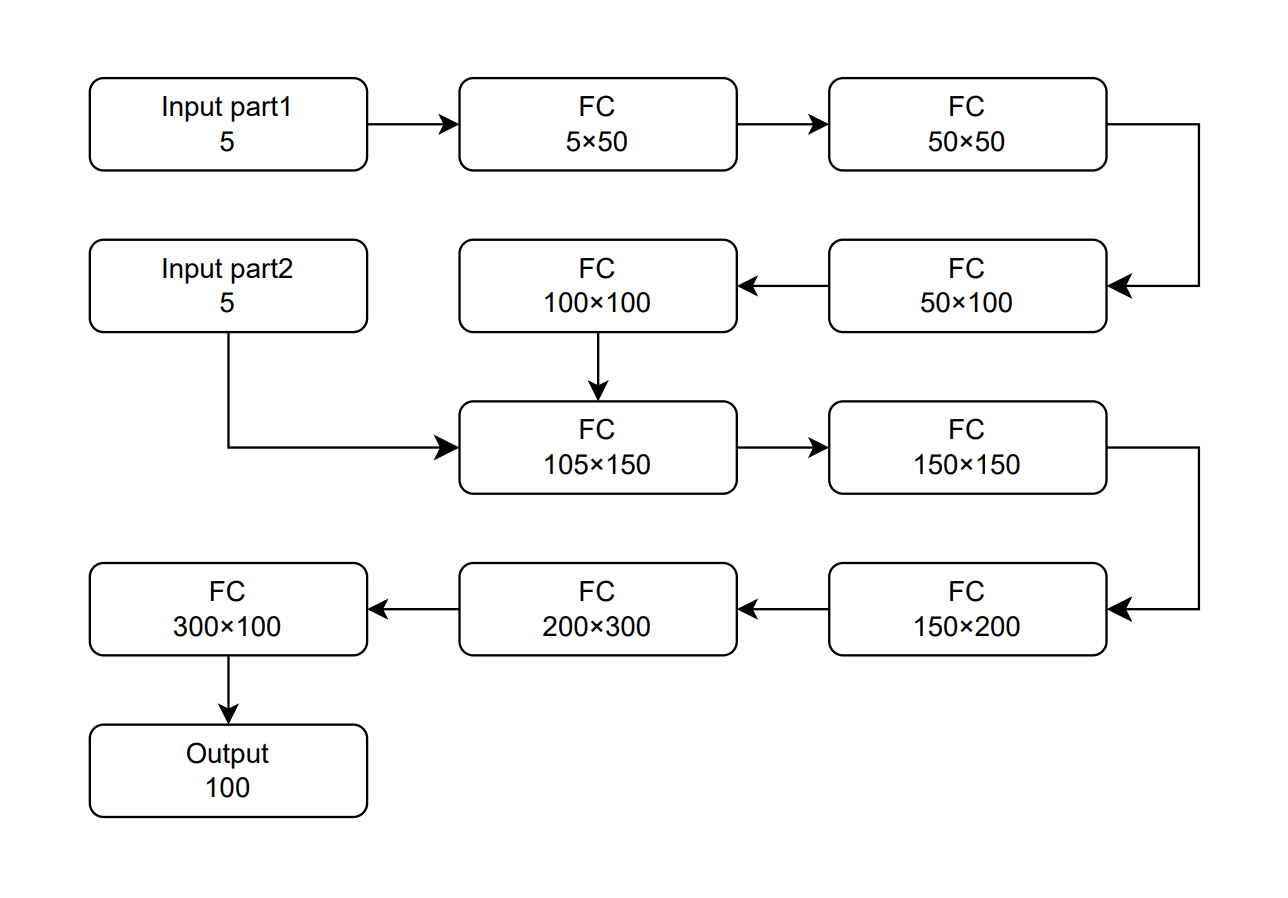}
\caption{The structure of the network.}
\label{Fig1}
\end{figure*}

We first applied a scaling transformation to the parameters. This process helps standardize the parameter magnitudes, thus enhancing model performance during training: $T_1$/7000, $i$/90, $\theta/180$, $\lambda/360$, $r_s/60$, $l_3$/0.6. We employed a fully connected NN for training. Since the results obtained by directly inputting all the 10 parameters for training were not satisfactory, we designed a network structure.
The original 10-dimensional input parameters are divided into two 5-dimensional subsets. The first subset contains [$T_1$, $T_2/T_1$, $q$, $i$, $f$], and the second one includes four starspot parameters plus third light. The structure of the network is shown in Figure \ref{Fig1}. We first input the five parameters of the first subset into the network. These parameters pass through a fully connected layer of size 5×50, followed by a 50×50 fully connected layer, a 50×100 fully connected layer, and a 100×100 fully connected layer. Subsequently, we concatenate the five parameters of the second subset with the 100-dimensional output from the previous layer, forming a 105-dimensional data. This 105-dimensional data then passes through a 105×150 fully connected layer, a 150×150 fully connected layer, a 150×200 fully connected layer, a 200×300 fully connected layer, and a 300×100 fully connected layer. The final 100-dimensional output represents the LC. 
Through the above process, we have obtained a NN model for the V-band LC. Subsequently, we trained another model using the same architecture for calculating luminosities of the two components (with the output dimension changed to two). Considering that the radii of the primary and secondary component stars, as well as the potential, are only related to $q$ and $f$, the input parameters only consist of these two parameters. The structure is a fully connected NN with a size of 3×100, and the output dimension is three. All activation functions used are ReLU functions \citep{2015arXiv150201852H}, and the optimizer used throughout is Adam \citep{2014arXiv1412.6980K}.
Thus, we have developed three models: one for generating LCs, one for generating luminosity values, and one for generating radii and potential.

To address the multiple-band problem while avoiding the need for large training datasets, we employed the transfer learning approach \citep{5288526}. By performing full-parameter fine-tuning based on the pre-trained V band model, this method allows us to train a well-performing model with a small dataset. The required training data size ranges from 5,000 to 50,000, as the shapes of LCs differ significantly across different bands compared to the V band. Accordingly, we developed the corresponding three NN models as V band for Johnson: BRI, Cousins: RI, ZTF: gri, SDSS: griz, TESS: T, Kepler: mean, Gaia: GBPRP, and SuperWASP: post2004 bands. The goodness-of-fit ($R^2$) between the predicted and PHOEBE generated LCs and the deviation in the primary luminosity ratio for each band in the test set are shown in Table \ref{tab:test}. The radius and potential deviations are as follows, $r_1$: mean value is 0.0003, median value is 0.0001, $r_2$: mean value is 0.0003, median value is 0.0001, potential: mean value is 0.0003, median value is 0.0001.

\begin{table}
\centering
\small
\caption{ The results of the test set for each band of our model} \label{tab:test}
\begin{tabular}{lcccc}
\hline
Filter	            &LC (mean R$^2$) 	&LC (median R$^2$) 	&L$_2$/L$_T$ deviation (mean)	&L$_2$/L$_T$ deviation (median) \\ \hline
Johnson: B	        &0.9990 	        &  0.9997 	        &     0.0010 	                &     0.0008                    \\
Johnson: V	        &0.9991 	        &  0.9997 	        &     0.0008 	                &     0.0007                    \\
Johnson: R	        &0.9991 	        &  0.9998 	        &     0.0008 	                &     0.0006                    \\
Johnson: I	        &0.9991 	        &  0.9998 	        &     0.0009 	                &     0.0007                    \\
Cousins: R	        &0.9990 	        &  0.9997 	        &     0.0009 	                &     0.0007                    \\
Cousins: I	        &0.9990 	        &  0.9998 	        &     0.0009 	                &     0.0007                    \\
ZTF: g	            &0.9990 	        &  0.9997 	        &     0.0008 	                &     0.0006                    \\
ZTF: r	            &0.9992 	        &  0.9998 	        &     0.0007 	                &     0.0006                    \\
ZTF: i	            &0.9991 	        &  0.9998 	        &     0.0009 	                &     0.0007                    \\
SDSS: g	            &0.9990 	        &  0.9997 	        &     0.0008 	                &     0.0006                    \\
SDSS: r	            &0.9990 	        &  0.9998 	        &     0.0007 	                &     0.0006                    \\
SDSS: i	            &0.9992 	        &  0.9998 	        &     0.0008 	                &     0.0007                    \\
SDSS: z	            &0.9991 	        &  0.9998 	        &     0.0010 	                &     0.0008                    \\
TESS: T	            &0.9991 	        &  0.9997 	        &     0.0008 	                &     0.0006                    \\
Kepler: mean	      &0.9992 	        &  0.9998 	        &     0.0007 	                &     0.0006                    \\
Gaia: G	            &0.9991 	        &  0.9997 	        &     0.0007 	                &     0.0006                    \\
Gaia: BP	          &0.9990 	        &  0.9997 	        &     0.0008 	                &     0.0007                    \\
Gaia: RP	          &0.9991 	        &  0.9998 	        &     0.0008 	                &     0.0007                    \\
SuperWASP: post2004	&0.9990 	        &  0.9997 	        &     0.0007 	                &     0.0006                    \\
\hline
\end{tabular}
\end{table}                               

\section{Physical parameter determination and the robustness of our model} 
To determine the physical parameters of contact binaries from their LCs using our built NN models, we implemented a two-stage MCMC process: (1) In the first stage, we performed a MCMC calculation with 64 walkers and 2000 iterations. The range of parameters follows that used in the training of NN model and is uniform in distribution. (2) In the second stage, we adopted the parameter values from the chain with the highest average likelihood in the first stage as our priors. These priors were assigned Gaussian distributions, with their uncertainties set to the range of the corresponding Gaussian distributions.
After the two-stage MCMC process, physical parameters of a contact binary can be obtained. The computations were performed on a computer equipped with 64 GB of RAM and a multi-core CPU configuration, consisting of an 8-core processor running at 3.2 GHz and a 16-core processor operating at 2.4 GHz. For LCs of a contact binary with three bands and comprising 729 data points in total for all bands, the processing time was 82 s. In contrast, the same analysis required 4.8 days when performed using PHOEBE.

\begin{figure*}[htbp]
\includegraphics[width=9cm]{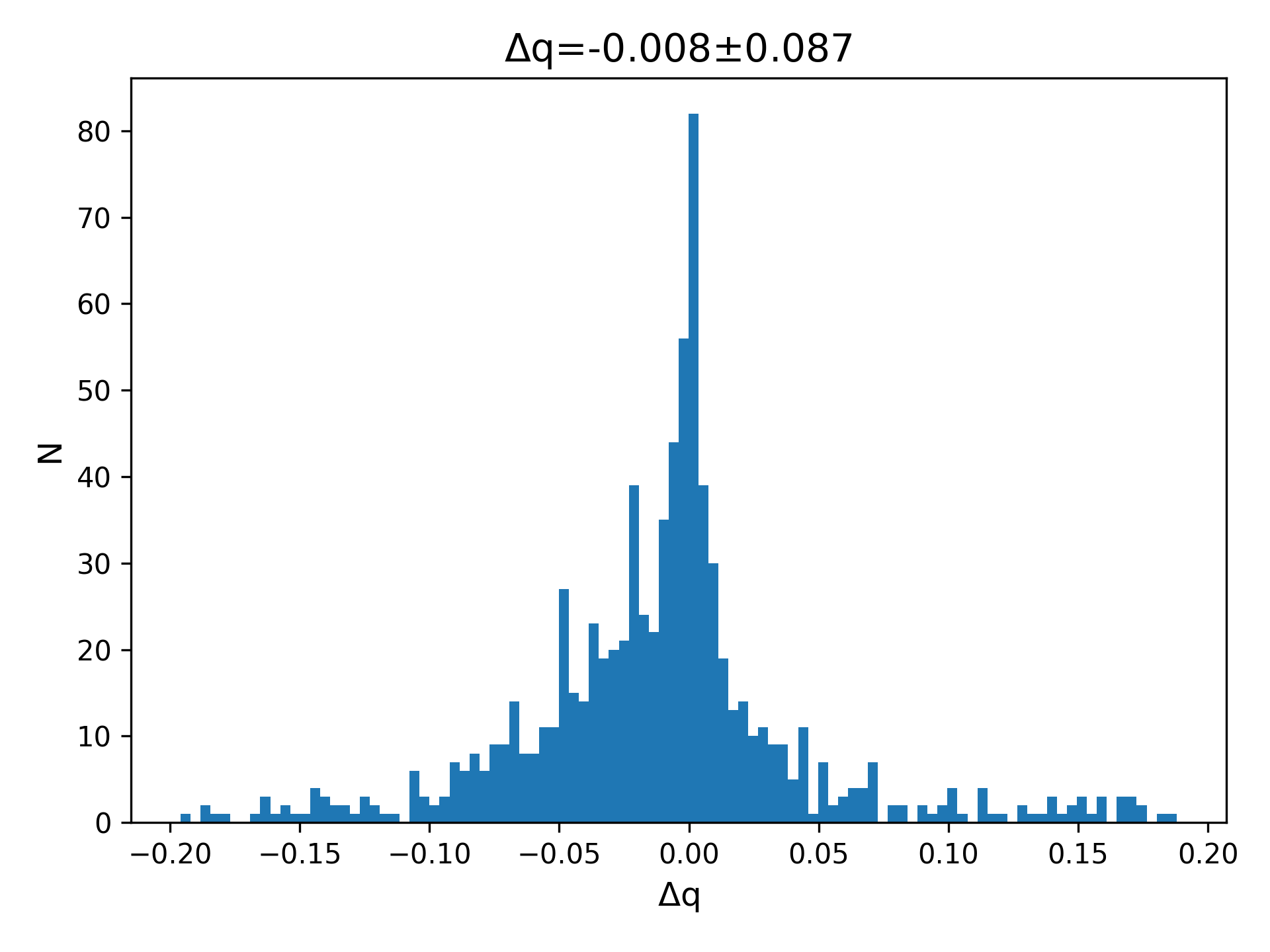}
\includegraphics[width=9cm]{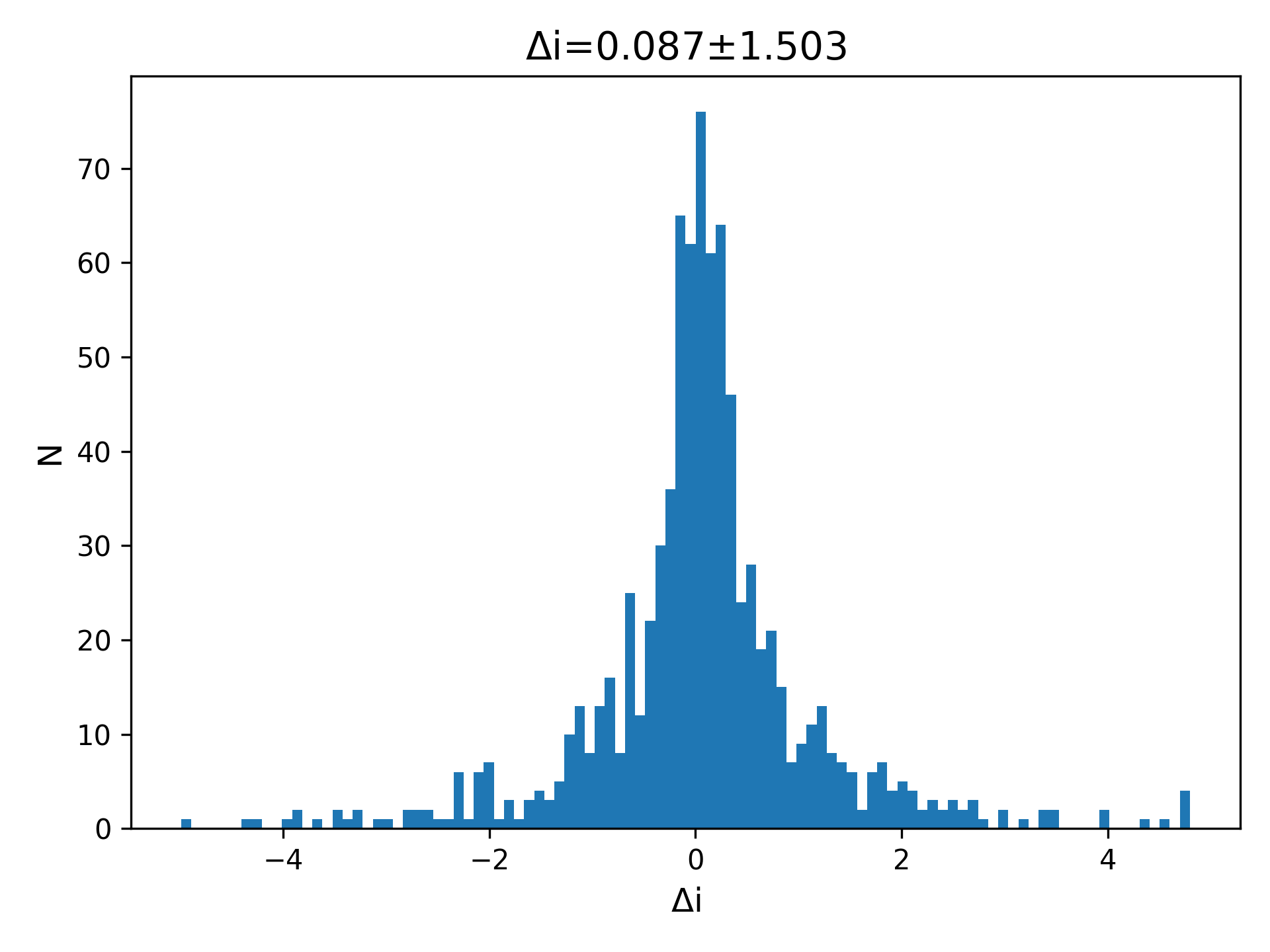}
\includegraphics[width=9cm]{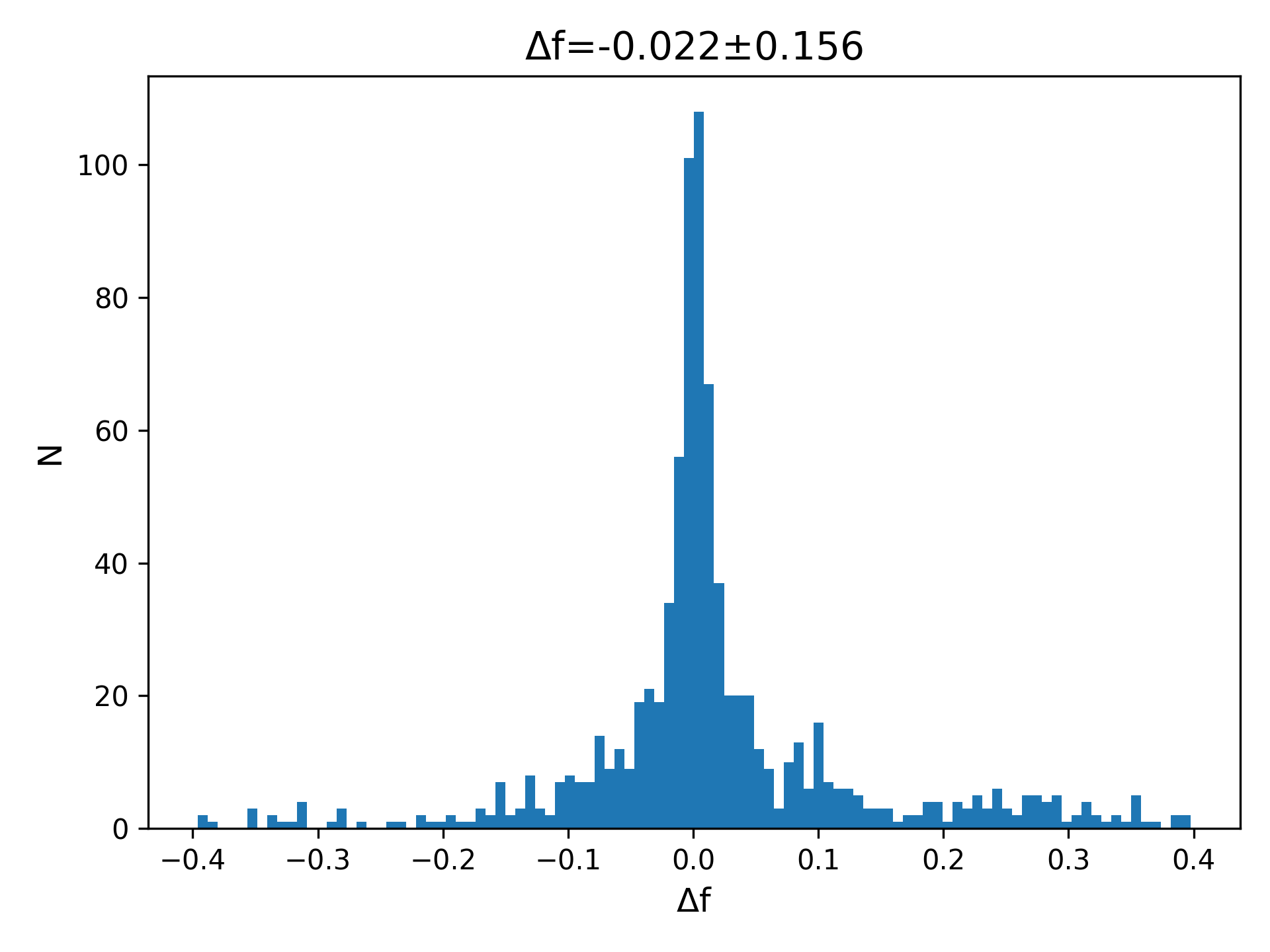}
\includegraphics[width=9cm]{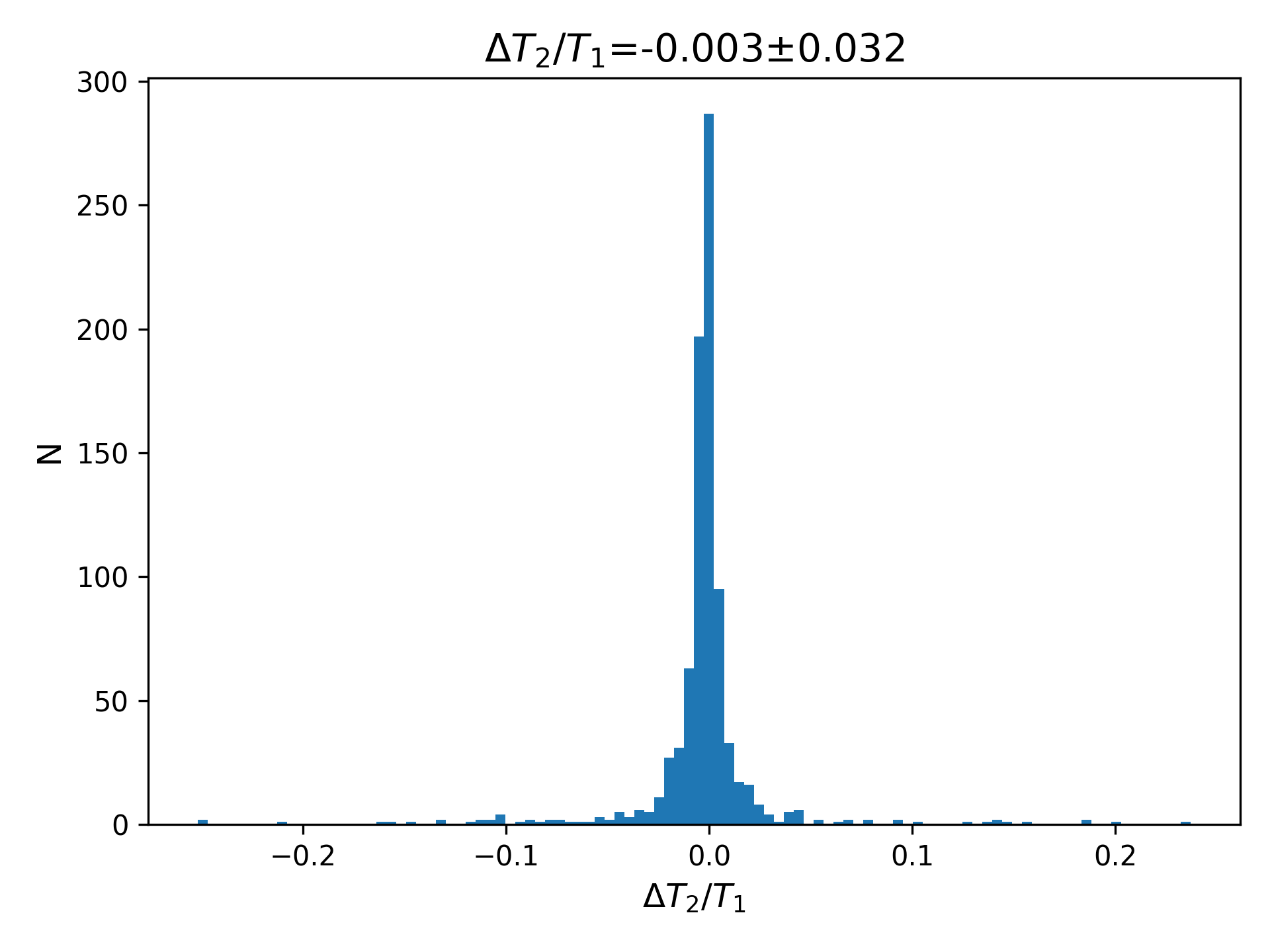}
\includegraphics[width=9cm]{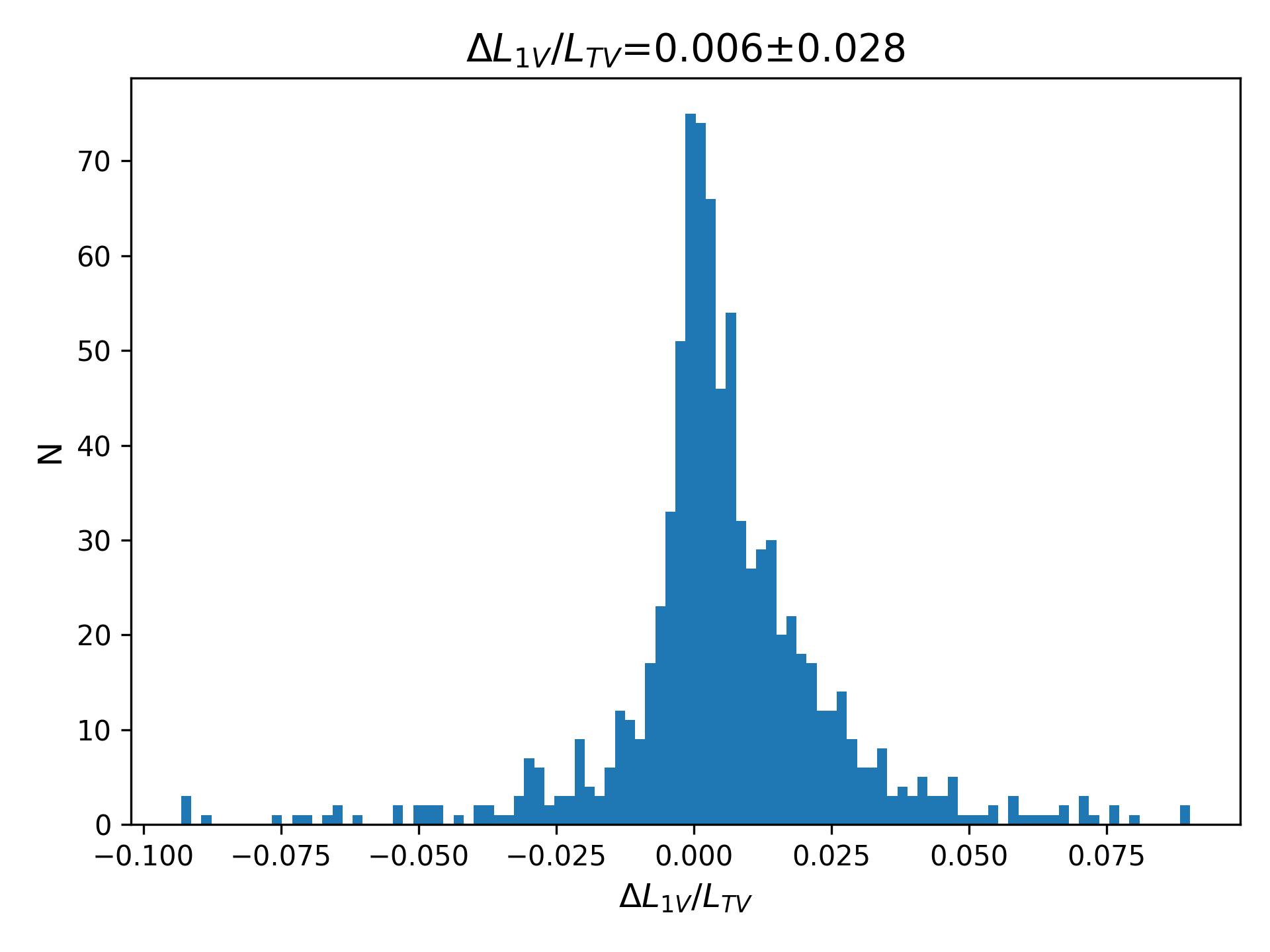}
\includegraphics[width=9cm]{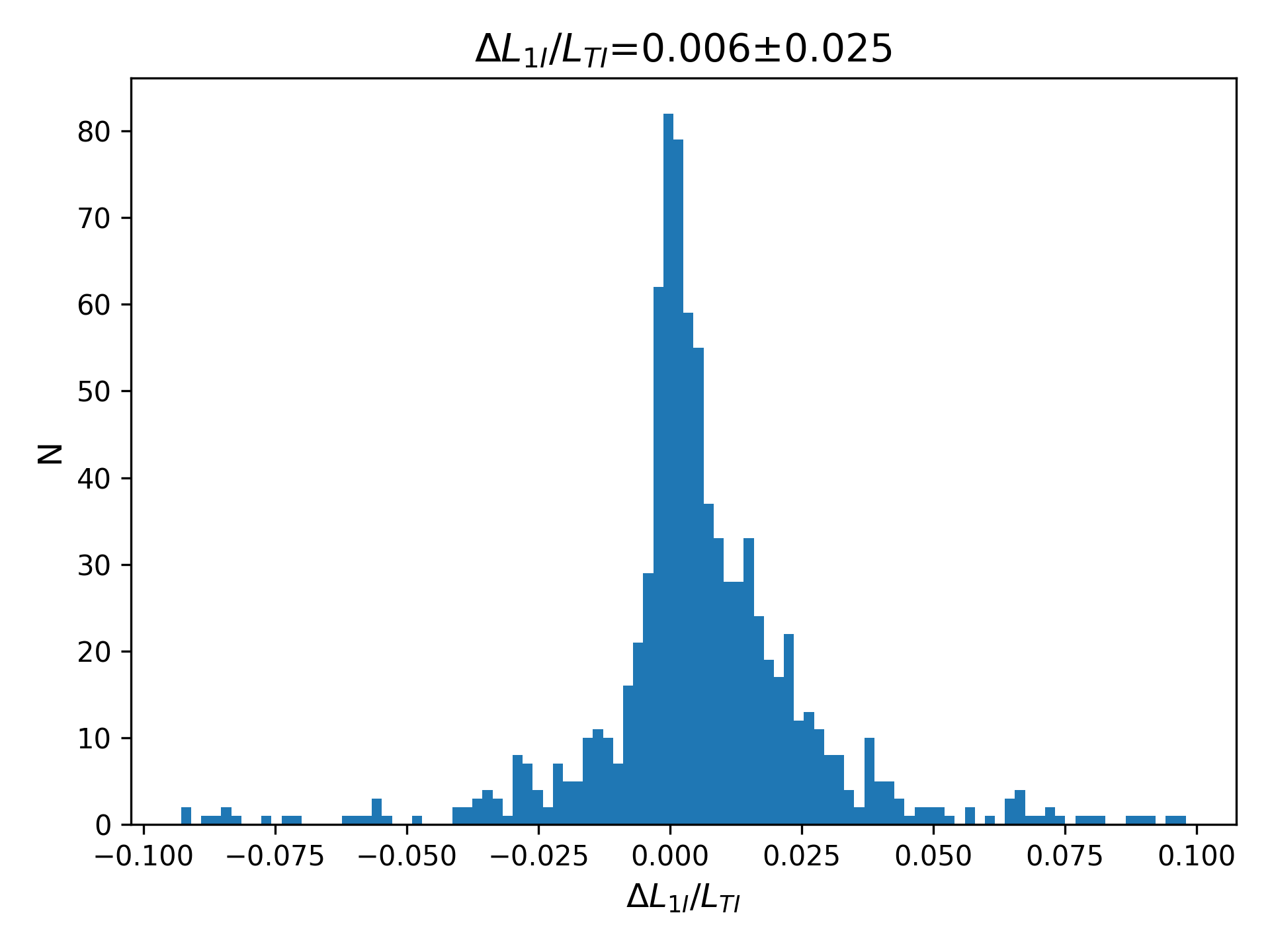}
\caption{The distribution of the difference between the physical parameters obtained.}
\label{fig2}
\end{figure*}

\begin{figure*}[htbp]
\includegraphics[width=9cm]{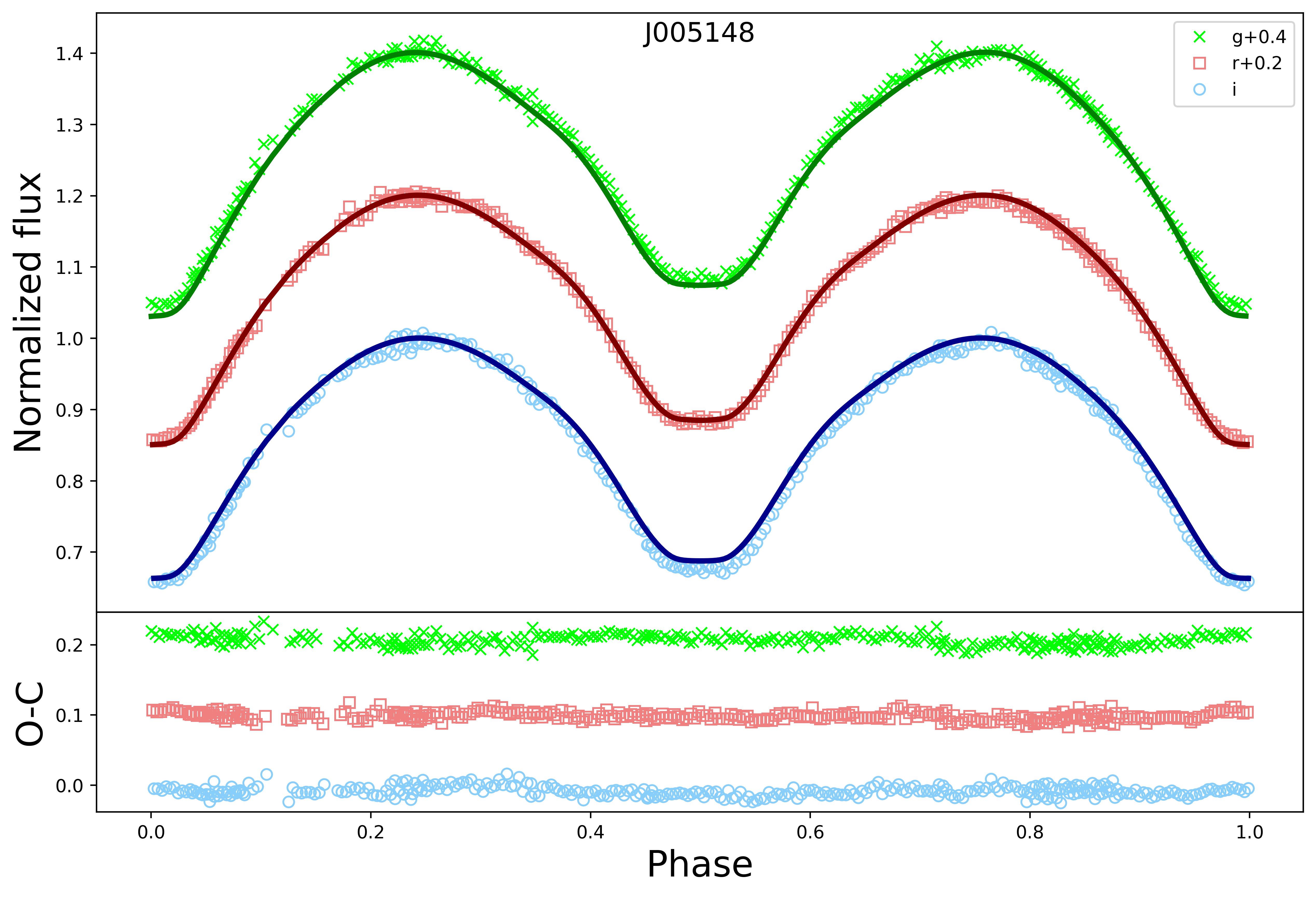}
\includegraphics[width=9cm]{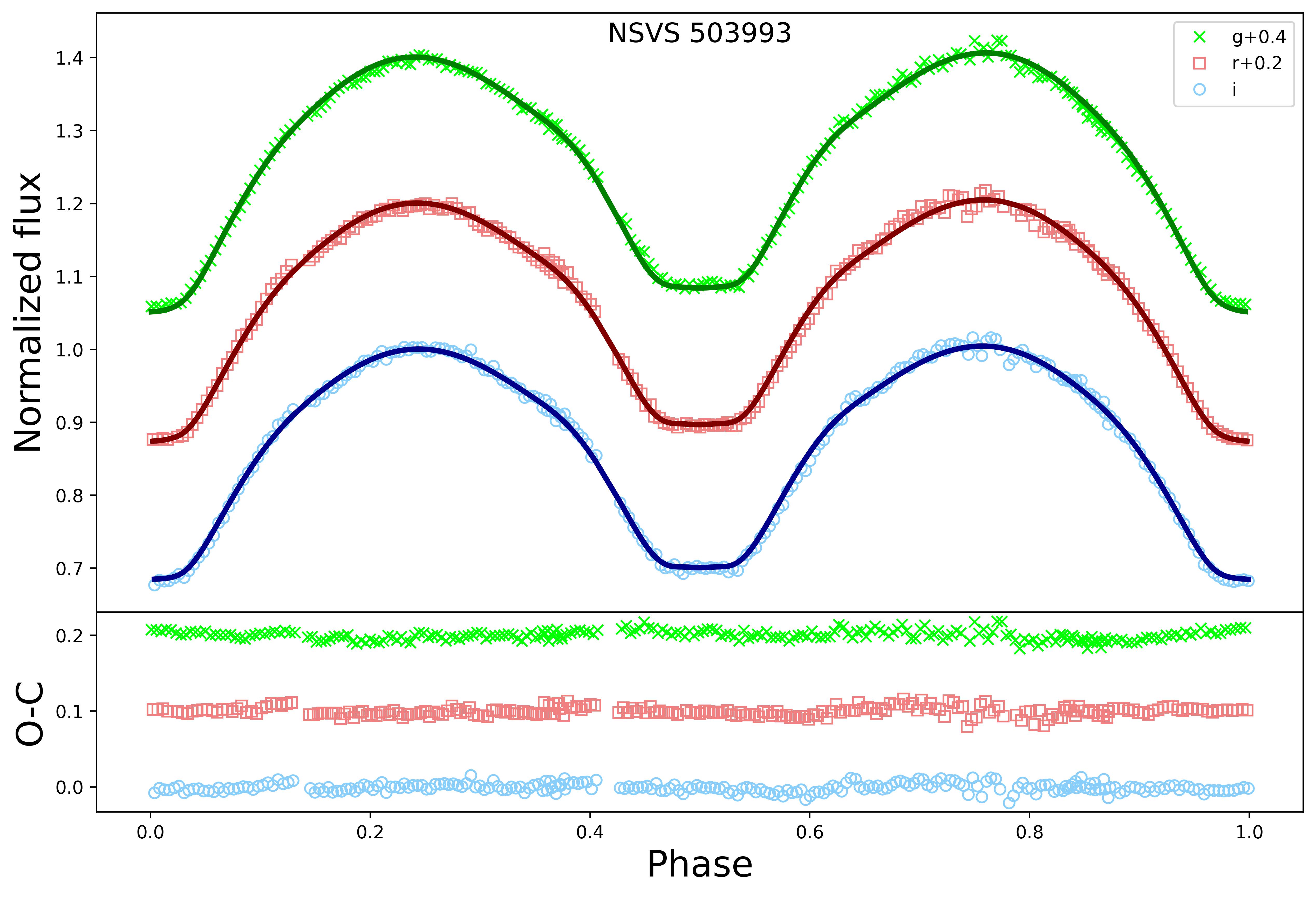}
\includegraphics[width=9cm]{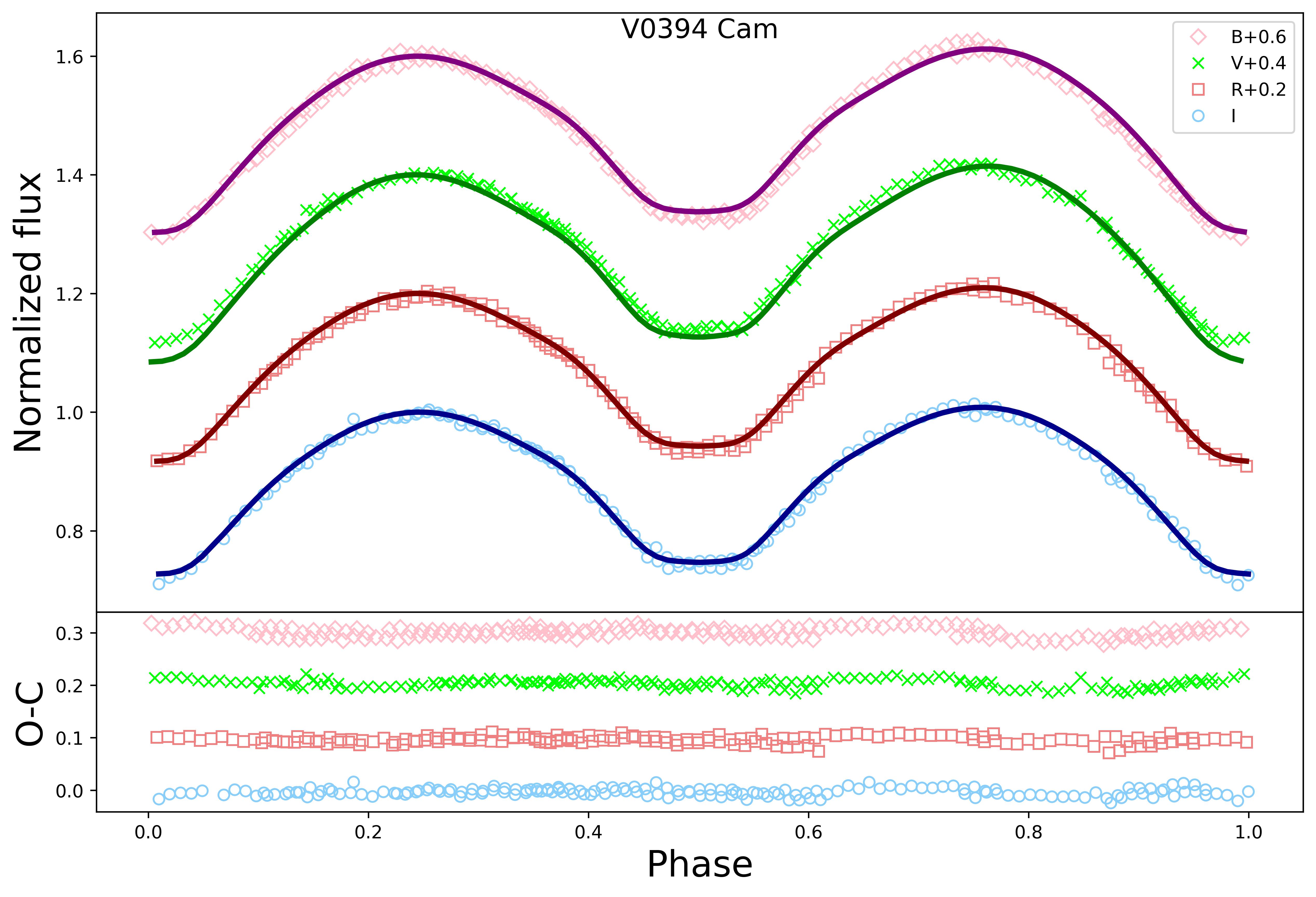}
\includegraphics[width=9cm]{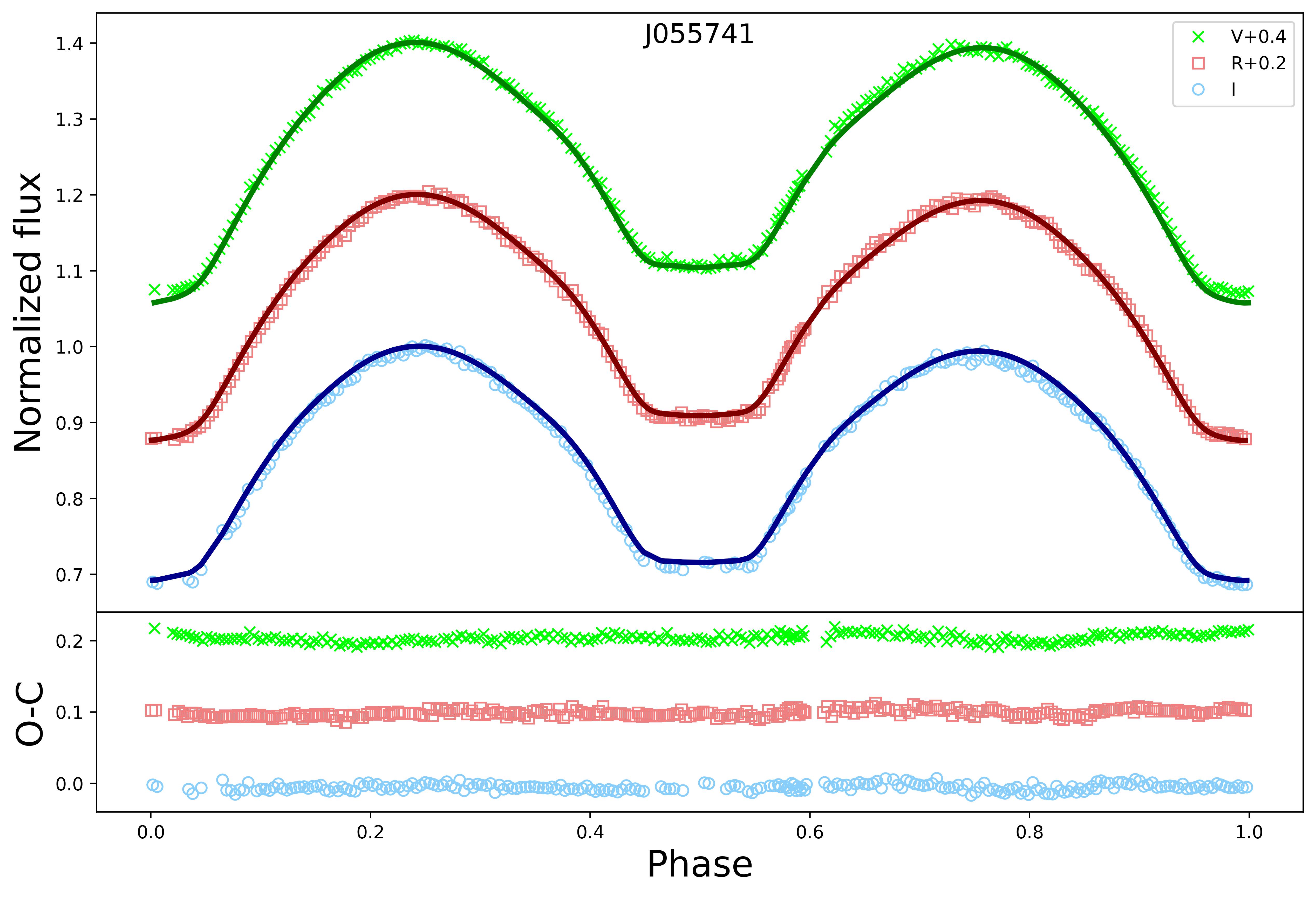}
\includegraphics[width=9cm]{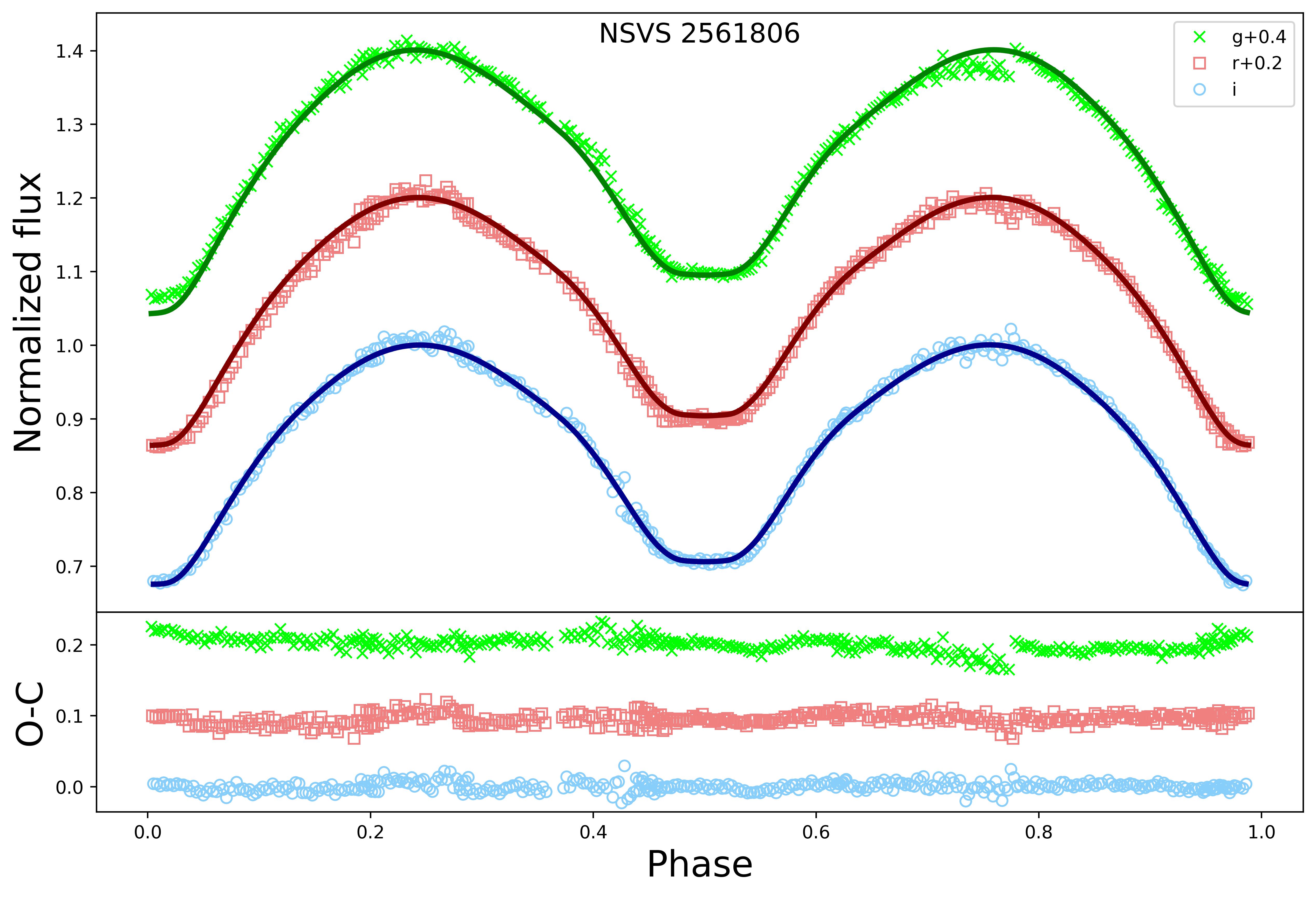}
\includegraphics[width=9cm]{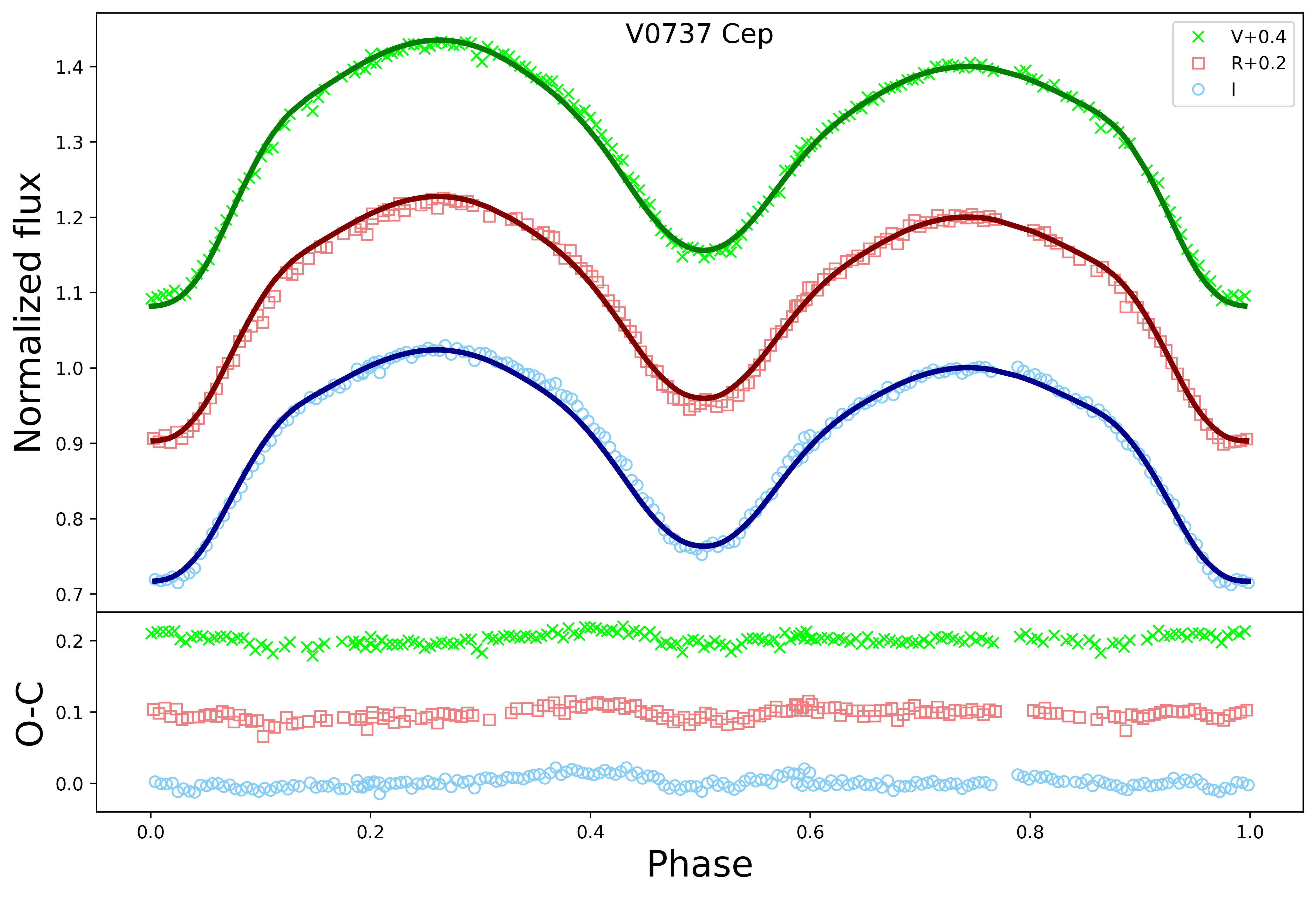}
\includegraphics[width=9cm]{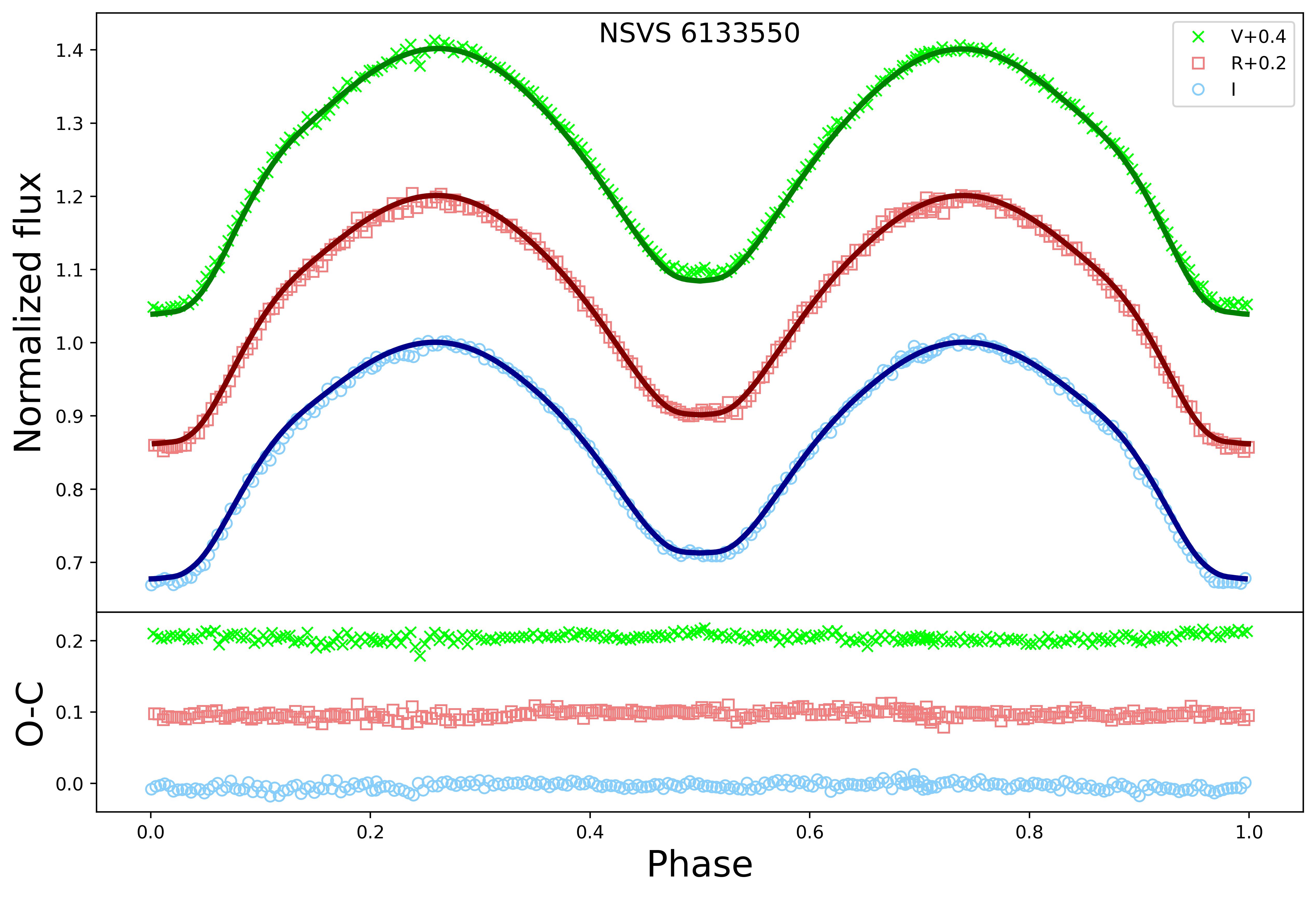}
\includegraphics[width=9cm]{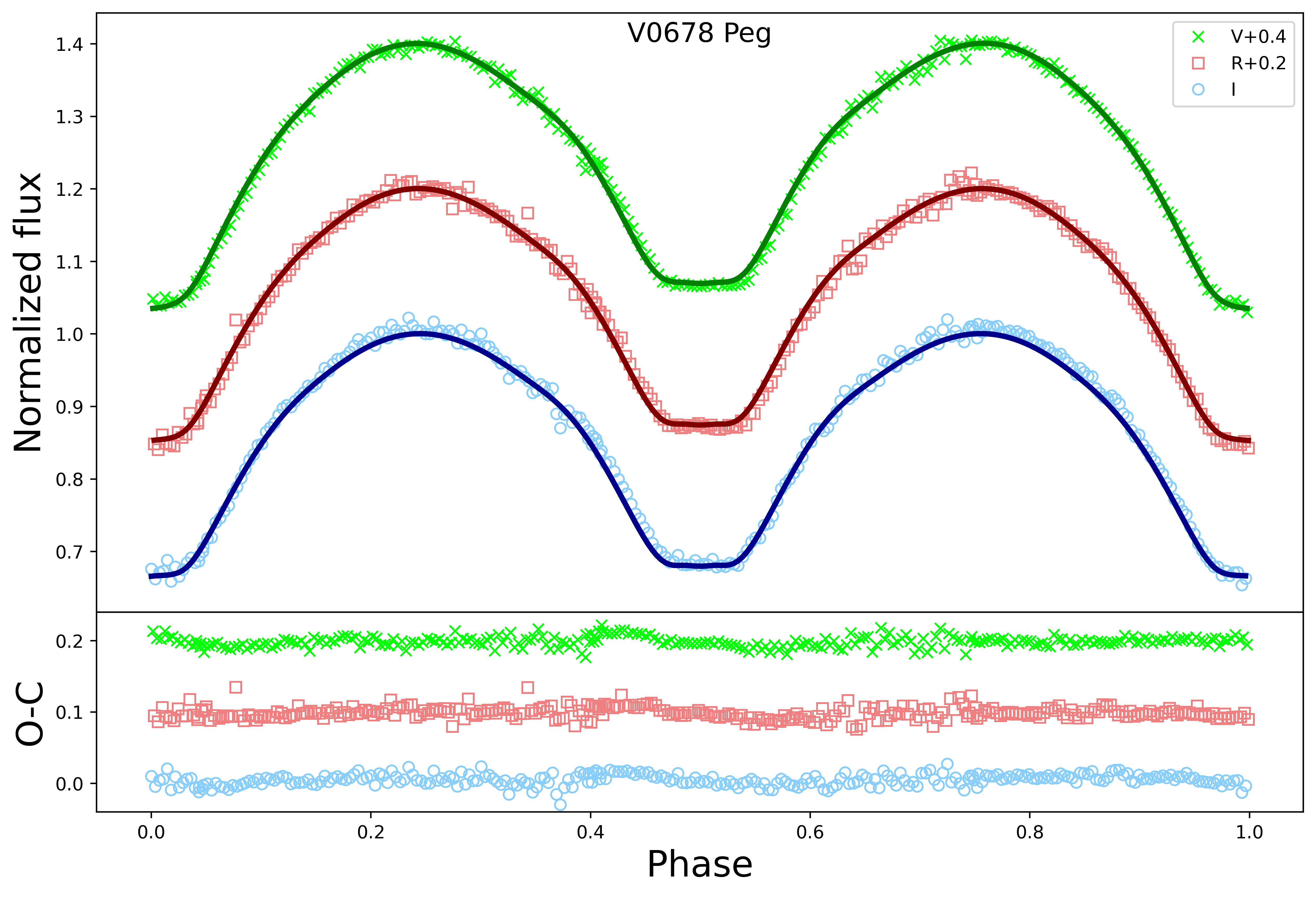}
\caption{The fitted LCs of the eight targets.}
\label{fig3}
\end{figure*}

To validate the robustness of our model, we applied it to two distinct datasets for comparison, 1000 sets of V and I synthetic LCs randomly generated by PHOEBE without third light or spot and LCs of eight targets studied by \cite{2024ApJ...976..223W}. Our NN model, implemented with the two-stage MCMC approach, successfully analyzed these LCs and determined their physical parameters. The discrepancy distribution between our model-derived physical parameters and and the true values of the 1000 sets of synthetic LCs is shown in Figure \ref{fig2}. These parameters exhibit very good consistency. The derived physical parameters of the eight targets from \cite{2024ApJ...976..223W} are presented in Table \ref{tab:PP}. The corresponding results from \cite{2024ApJ...976..223W} are also included in this table for direct comparison, revealing remarkable consistency between the two sets of physical parameters. The comparison between theoretical and observational LCs are displayed in Figure \ref{fig3}. The fitted LCs also exhibit excellent consistency with the results reported by \cite{2024ApJ...976..223W}. Strong agreement with the results reported by \cite{2024ApJ...976..223W}, along with the high-quality fits obtained for all targets, robustly validates both the reliability and accuracy of our model.

\begin{table}
\centering
\footnotesize
\caption{Comparison between the parameters determined by us and \cite{2024ApJ...976..223W} (Wang2024)} \label{tab:PP}
\begin{tabular}{lccccccccc}
\hline
Star                          &          &J005148                  &NSVS 503993              &V0394 Cam                &J055741                  &NSVS 2561806             &V0737 Cep                &NSVS 6133550             &V0678 Peg                \\\hline 
q              &This paper&0.241$^{+0.000}_{-0.000}$&0.222$^{+0.000}_{-0.000}$&0.158$^{+0.000}_{-0.000}$&0.182$^{+0.000}_{-0.001}$&0.217$^{+0.000}_{-0.001}$&0.170$^{+0.000}_{-0.000}$&0.174$^{+0.000}_{-0.000}$&0.245$^{+0.000}_{-0.000}$\\ 
                              &Wang2024  &0.237$^{+0.001}_{-0.001}$&0.220$^{+0.001}_{-0.001}$&0.157$^{+0.001}_{-0.001}$&0.188$^{+0.001}_{-0.001}$&0.219$^{+0.002}_{-0.002}$&0.160$^{+0.000}_{-0.000}$&0.174$^{+0.002}_{-0.002}$&0.250$^{+0.002}_{-0.002}$\\ 
i (deg)         &This paper&78.4$^{+0.0}_{-0.0}$     &80.4$^{+0.0}_{-0.0}$     &75.7$^{+0.0}_{-0.0}$     &82.4$^{+0.0}_{-0.0}$     &78.0$^{+0.0}_{-0.0}$     &73.6$^{+0.0}_{-0.0}$     &76.4$^{+0.0}_{-0.0}$     &81.9$^{+0.0}_{-0.0}$     \\ 
                              &Wang2024  &78.8$^{+0.2}_{-0.2}$     &80.1$^{+0.3}_{-0.3}$     &75.0$^{+0.2}_{-0.2}$     &83.9$^{+0.5}_{-0.4}$     &78.2$^{+0.2}_{-0.2}$     &73.6$^{+0.1}_{-0.2}$     &76.0$^{+0.2}_{-0.2}$     &81.5$^{+0.4}_{-0.3}$     \\ 
T$_2$ (K)       &This paper&6451$^{+1}_{-2}$         &6298$^{+1}_{-1}$         &5961$^{+1}_{-3}$         &6511$^{+2}_{-3}$         &6173$^{+0}_{-0}$         &4651$^{+1}_{-1}$         &5007$^{+1}_{-1}$         &6224$^{+1}_{-1}$         \\ 
                              &Wang2024  &6425$^{+24}_{-24}$       &6303$^{+38}_{-36}$       &6050$^{+33}_{-41}$       &6470$^{+27}_{-27}$       &6142$^{+27}_{-25}$       &4722$^{+23}_{-22}$       &5051$^{+21}_{-21}$       &6180$^{+27}_{-29}$       \\ 
$\Omega$        &This paper&2.283$^{+0.000}_{-0.000}$&2.259$^{+0.000}_{-0.000}$&2.077$^{+0.001}_{-0.000}$&2.120$^{+0.002}_{-0.001}$&2.223$^{+0.000}_{-0.000}$&2.151$^{+0.001}_{-0.001}$&2.114$^{+0.000}_{-0.000}$&2.307$^{+0.000}_{-0.000}$\\ 
                              &Wang2024  &2.278$^{+0.003}_{-0.003}$&2.255$^{+0.004}_{-0.003}$&2.083$^{+0.005}_{-0.003}$&2.146$^{+0.003}_{-0.005}$&2.234$^{+0.004}_{-0.006}$&2.125$^{+0.001}_{-0.001}$&2.117$^{+0.004}_{-0.004}$&2.315$^{+0.006}_{-0.006}$\\ 
$f(\%)$         &This paper&31.8$^{+0.0}_{-0.0}$     &19.9$^{+0.1}_{-0.1}$     &46.8$^{+0.1}_{-0.2}$     &57.0$^{+0.7}_{-0.6}$     &36.3$^{+0.2}_{-0.1}$     &5.1$^{+0.2}_{-0.1}$       &48.1$^{+0.1}_{-0.1}$     &22.1$^{+0.1}_{-0.1}$     \\ 
                              &Wang2024  &29.6$^{+0.9}_{-1.8}$     &18.5$^{+1.1}_{-0.9}$     &38.2$^{+1.5}_{-0.7}$     &46.3$^{+2.0}_{-1.3}$     &33.3$^{+1.1}_{-1.2}$     &36.0$^{+0.1}_{-0.1}$     &43.1$^{+1.3}_{-1.0}$     &23.8$^{+1.3}_{-1.4}$     \\ 
r$_1$           &This paper&0.523$^{+0.000}_{-0.000}$&0.523$^{+0.000}_{-0.000}$&0.561$^{+0.000}_{-0.000}$&0.555$^{+0.000}_{-0.001}$&0.533$^{+0.000}_{-0.001}$&0.540$^{+0.000}_{-0.000}$&0.555$^{+0.000}_{-0.000}$&0.517$^{+0.000}_{-0.000}$\\ 
                              &Wang2024  &0.523$^{+0.001}_{-0.001}$&0.524$^{+0.001}_{-0.001}$&0.559$^{+0.001}_{-0.001}$&0.549$^{+0.001}_{-0.001}$&0.531$^{+0.001}_{-0.001}$&0.544$^{+0.001}_{-0.001}$&0.553$^{+0.001}_{-0.001}$&0.516$^{+0.001}_{-0.001}$\\ 
r$_2$           &This paper&0.282$^{+0.000}_{-0.000}$&0.269$^{+0.000}_{-0.000}$&0.256$^{+0.000}_{-0.000}$&0.272$^{+0.000}_{-0.000}$&0.275$^{+0.000}_{-0.000}$&0.242$^{+0.000}_{-0.000}$&0.264$^{+0.000}_{-0.000}$&0.278$^{+0.000}_{-0.000}$\\ 
                              &Wang2024  &0.279$^{+0.001}_{-0.001}$&0.267$^{+0.001}_{-0.001}$&0.251$^{+0.001}_{-0.001}$&0.269$^{+0.001}_{-0.001}$&0.275$^{+0.001}_{-0.001}$&0.237$^{+0.001}_{-0.001}$&0.261$^{+0.001}_{-0.001}$&0.280$^{+0.001}_{-0.001}$\\ 
$L_{1g}/L_{Tg}$ &This paper&0.790$^{+0.000}_{-0.000}$&0.790$^{+0.000}_{-0.000}$&$-$                      &$-$                      &0.809$^{+0.000}_{-0.000}$&$-$                      &$-$                      &$-$                      \\               
                              &Wang2024  &0.796$^{+0.001}_{-0.001}$&0.793$^{+0.002}_{-0.002}$&$-$                      &$-$                      &0.814$^{+0.001}_{-0.001}$&$-$                      &$-$                      &$-$                      \\               
$L_{1r}/L_{Tr}$ &This paper&0.787$^{+0.000}_{-0.000}$&0.793$^{+0.000}_{-0.000}$&$-$                      &$-$                      &0.806$^{+0.000}_{-0.000}$&$-$                      &$-$                      &$-$                      \\               
                              &Wang2024  &0.792$^{+0.001}_{-0.001}$&0.794$^{+0.001}_{-0.001}$&$-$                      &$-$                      &0.808$^{+0.001}_{-0.001}$&$-$                      &$-$                      &$-$                      \\               
$L_{1i}/L_{Ti}$ &This paper&0.786$^{+0.000}_{-0.000}$&0.792$^{+0.000}_{-0.000}$&$-$                      &$-$                      &0.804$^{+0.000}_{-0.000}$&$-$                      &$-$                      &$-$                      \\               
                              &Wang2024  &0.790$^{+0.001}_{-0.001}$&0.794$^{+0.001}_{-0.001}$&$-$                      &$-$                      &0.805$^{+0.001}_{-0.001}$&$-$                      &$-$                      &$-$                      \\               
$L_{1B}/L_{TB}$ &This paper&$-$                      &$-$                      &0.833$^{+0.000}_{-0.000}$&$-$                      &$-$                      &$-$                      &$-$                      &$-$                      \\                   
                              &Wang2024  &$-$                      &$-$                      &0.825$^{+0.001}_{-0.001}$&$-$                      &$-$                      &$-$                      &$-$                      &$-$                      \\                   
$L_{1V}/L_{TV}$ &This paper&$-$                      &$-$                      &0.835$^{+0.000}_{-0.000}$&0.812$^{+0.000}_{-0.000}$&$-$                      &0.750$^{+0.000}_{-0.000}$&0.748$^{+0.000}_{-0.000}$&0.776$^{+0.000}_{-0.000}$\\        
                              &Wang2024  &$-$                      &$-$                      &0.828$^{+0.001}_{-0.001}$&0.815$^{+0.001}_{-0.001}$&$-$                      &0.761$^{+0.001}_{-0.001}$&0.749$^{+0.001}_{-0.001}$&0.778$^{+0.002}_{-0.002}$\\       
$L_{1R}/L_{TR}$ &This paper&$-$                      &$-$                      &0.835$^{+0.000}_{-0.000}$&0.813$^{+0.000}_{-0.000}$&$-$                      &0.769$^{+0.000}_{-0.000}$&0.764$^{+0.000}_{-0.000}$&0.777$^{+0.000}_{-0.000}$\\        
                              &Wang2024  &$-$                      &$-$                      &0.829$^{+0.001}_{-0.001}$&0.814$^{+0.001}_{-0.001}$&$-$                      &0.781$^{+0.001}_{-0.001}$&0.765$^{+0.001}_{-0.001}$&0.777$^{+0.002}_{-0.002}$\\        
$L_{1I}/L_{TI}$ &This paper&$-$                      &$-$                      &0.834$^{+0.000}_{-0.000}$&0.812$^{+0.000}_{-0.000}$&$-$                      &0.781$^{+0.000}_{-0.000}$&0.774$^{+0.000}_{-0.000}$&0.778$^{+0.000}_{-0.000}$\\        
                              &Wang2024  &$-$                      &$-$                      &0.830$^{+0.001}_{-0.001}$&0.813$^{+0.001}_{-0.001}$&$-$                      &0.793$^{+0.001}_{-0.001}$&0.775$^{+0.001}_{-0.001}$&0.777$^{+0.002}_{-0.002}$\\        
$\lambda$ (deg) &This paper&$-$                      &284$^{+0}_{-0}$          &317$^{+0}_{-0}$          &63$^{+1}_{-10}$          &$-$                      &282$^{+0}_{-0}$          &$-$                      &$-$                      \\           
                              &Wang2024  &$-$                      &287$^{+4}_{-4}$          &351$^{+1}_{-1}$          &10$^{+1}_{-1}$           &$-$                      &287$^{+2}_{-2}$          &$-$                      &$-$                      \\           
$r_s$ (deg)     &This paper&$-$                      &7$^{+0}_{-0}$            &11$^{+0}_{-0}$           &5$^{+0}_{-0}$            &$-$                      &17$^{+0}_{-0}$           &$-$                      &$-$                      \\           
                              &Wang2024  &$-$                      &7$^{+1}_{-1}$            &25$^{+1}_{-1}$           &8$^{+1}_{-1}$            &$-$                      &22$^{+1}_{-1}$           &$-$                      &$-$                      \\           
$T_s$           &This paper&$-$                      &0.88$^{+0.00}_{-0.00}$   &0.89$^{+0.00}_{-0.00}$   &0.63$^{+0.02}_{-0.01}$   &$-$                      &0.90$^{+0.00}_{-0.00}$   &$-$                      &$-$                      \\           
                              &Wang2024  &$-$                      &0.91$^{+0.03}_{-0.04}$   &0.94$^{+0.01}_{-0.01}$   &0.82$^{+0.02}_{-0.03}$   &$-$                      &0.96$^{+0.00}_{-0.00}$   &$-$                      &$-$                      \\           
\hline
\end{tabular}
\end{table}                                         
\section{The application of our NN model} 
To demonstrate our model's capability in analyzing multiple-band data and its processing speed, we applied it to the data of Optical Gravitational Lensing Experiment (OGLE, \citealt{1994ApJ...426L..69U,2015AcA....65....1U}). Based on the study of \cite{2016AcA....66..405S},  86,560 contact binaries were identified, and data for their V and I bands are available for public download. We download the data of these binaries from OGLE-IV database at \url{ftp://ftp.astrouw.edu.pl/ogle/ogle4/OCVS/blg/ecl/}. In order to derive reliable physical parameters, we adopted the following two criteria for target selection: first, each target was required to have more than 100 data points in both the V and I bands; second, targets must possess cross-matched temperature information from Gaia DR3 \citep{2016A&A...595A...1G,2023A&A...674A...1G}. We obtained 5,332 targets, and for the subsequent analysis, data points below the 1st percentile and above the 99th percentile were removed from this sample. Using the orbital period and the time of eclipsing minimum determined by \cite{2016AcA....66..405S}, we converted the time to orbital phase and the magnitude measurements to flux. The resulting flux values were then normalized by setting the value at phase 0.25 to unity. To enhance computational efficiency, we binned the data points in each band into 100 points.

\begin{figure*}[htbp]
\includegraphics[width=9cm]{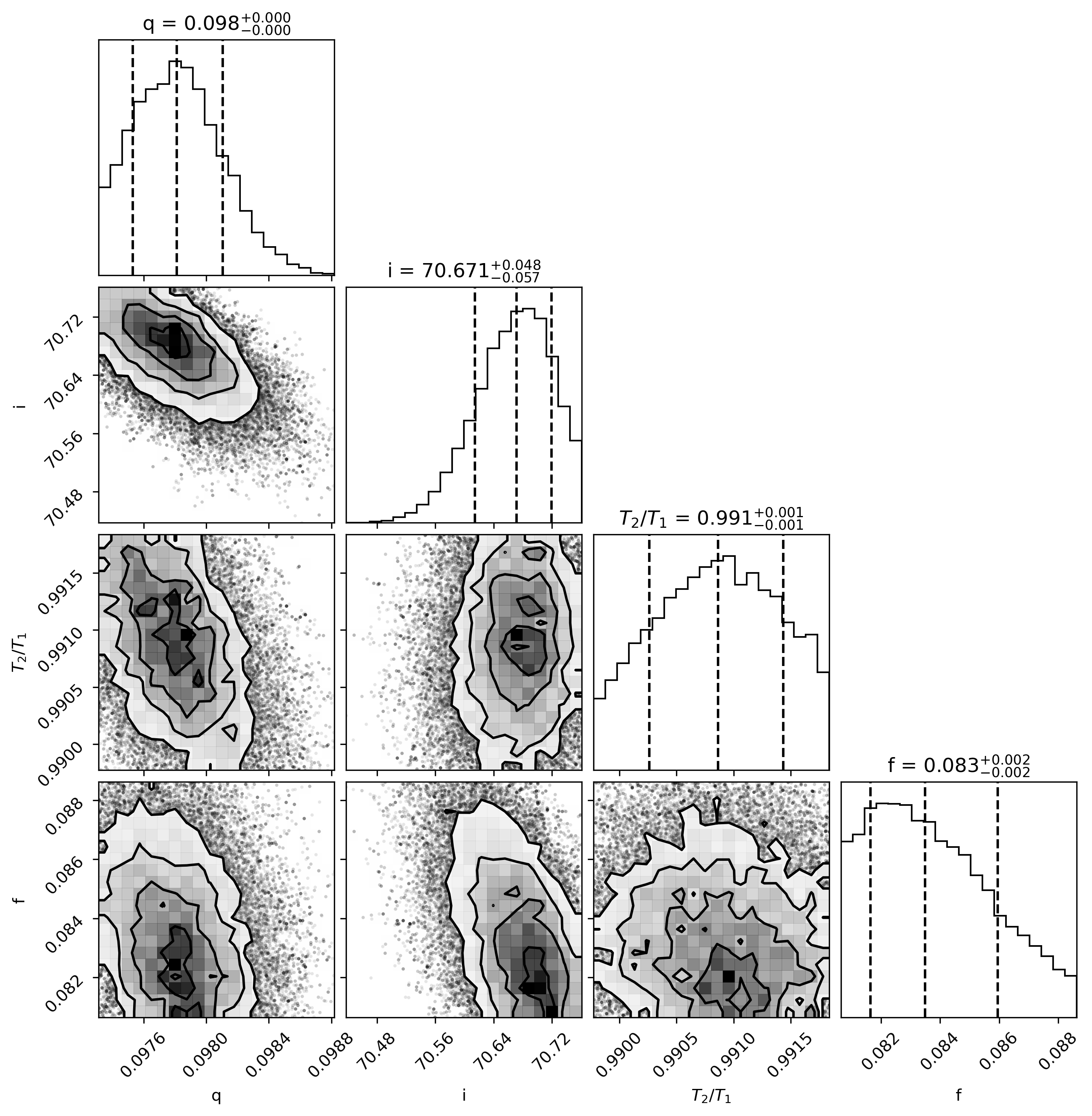}
\includegraphics[width=9cm]{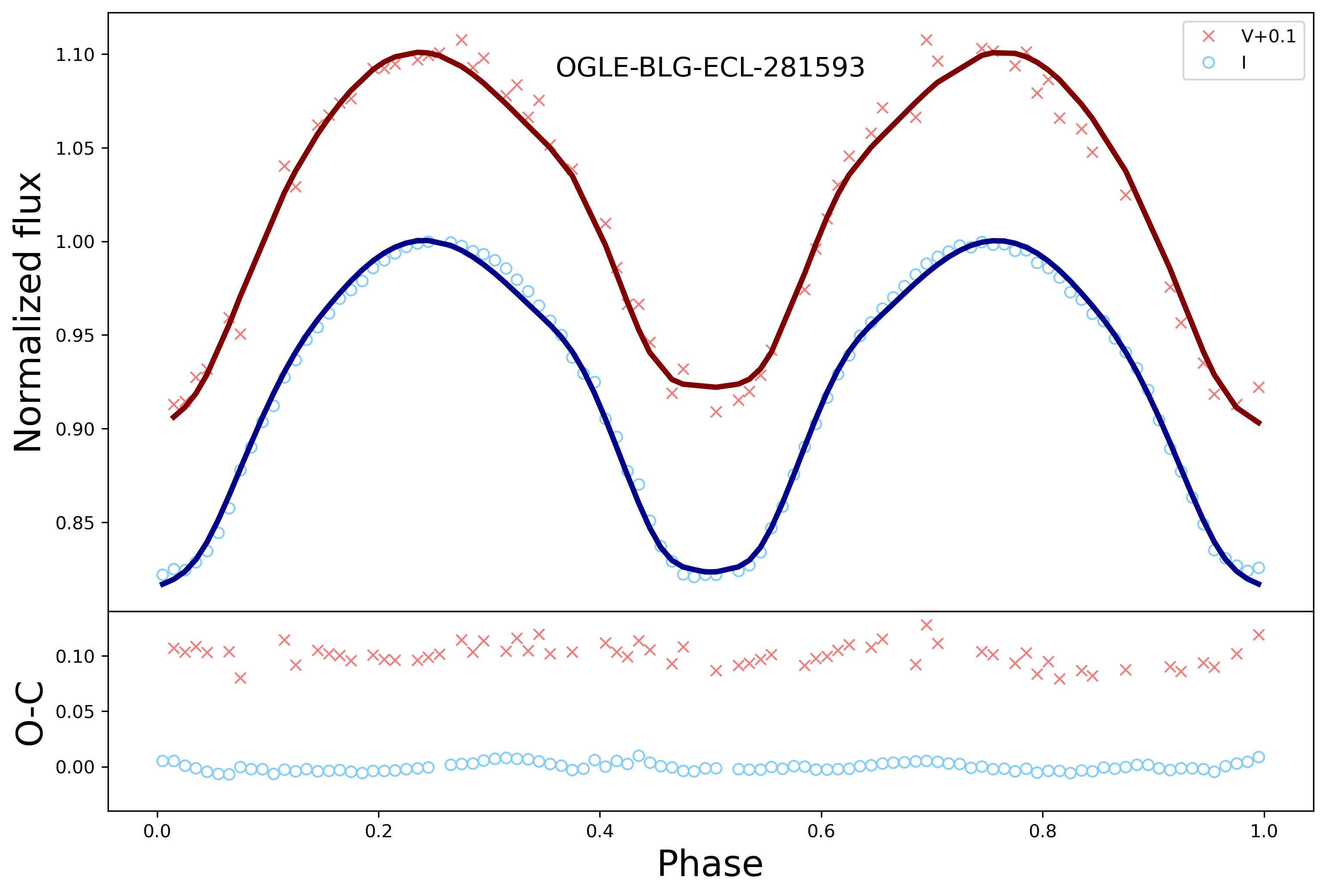}
\caption{The posterior parameter distributions and the best-fit model for an example system.}
\label{fig4}
\end{figure*}

During the analysis, spot parameters were included only if the brightness differences at the two maxima in the V and I bands both exceeded 0.01; otherwise, no spot parameters were considered. We analyzed each target four times using the NN model under the following conditions: phase shift = 0, without third light; phase shift = 0.5, without third light; phase shift = 0, with third light; phase shift = 0.5, with third light. The result with the highest $R^2$ value was selected as the final outcome for each target. After excluding systems with $R^2<0.8$, we derived physical parameters of a total of 3,541 targets. An example of the best-fitting model, along with the corresponding posterior parameter distributions, is presented in Figure \ref{fig4}. The derived physical parameters are provided in Table \ref{tab:Ph}.

\begin{table}
\small
\centering
\caption{ The physical parameters of the 3,541 systems} 
\begin{tabular}{llll}
\hline
Number & Column & Units & Explanation                             \\\hline
1      &  Name  &       & OGLE identifier                      \\
2      & $T_1$  & K     & Temperature of the primary component    \\
3      & q      &       & Mass ratio         \\
4      & $e_q$  &       & Uncertainty in q                        \\
5      & i      & deg   & Orbital inclination                     \\
6      & $e_i$  & deg   & Uncertainty in i                        \\
7      & $T_2/T_1$  &   & Temperature ratio                       \\  
8      & $e_{T_2/T_1}$& & Uncertainty in $T_2/T_1$                \\
9      & f             && Fillout factor                           \\
10     & $e_f$         && Uncertainty in f                         \\
11     & $\theta$      && Latitude of the spot                     \\
12     & $e_\theta$    && Uncertainty in $\theta$                  \\
13     & $\lambda$     && Longitude of the spot                    \\
14     & $e_\lambda$   && Uncertainty in $\lambda$                 \\
15     & $r_s$         && Angular radius of the spot               \\
16     & $e_{r_s}$     && Uncertainty in $r_s$                     \\
17     & $T_s$         && Relative temperature of the spot         \\
18     & $e_{T_s}$     && Uncertainty in $T_s$                     \\
19     &$L_2/L_{1V}$&      & Light ratio in V band                    \\
20     &$L_2/L_{1I}$&      & Light ratio in I band                 \\
21     &$l_3$&      & Third light ratio                        \\
22     &$e_{l_3}$&    & Uncertainty in $l_3$                \\
23     &$r_1$&     & Equival volume radius of the primary component  \\
24     &$r_2$&     & Equival volume radius of the secondary component \\
25     &$\Omega$&     & Dimensionless potential  \\
26     &$R^2$&     & Goodness of fit \\

\hline
\end{tabular}
\label{tab:Ph}
\end{table}

\section{Conclusion}                            
Based on the LCs generated by PHOEBE, we developed a NN model capable of rapidly analyzing multiple-band LCs of contact binaries and determining key physical parameters, including temperature ratio, mass ratio, orbital inclination, potential, contact degree, primary and secondary luminosities and radii, third light ratio, and four spot parameters. Unlike conventional approaches (e.g., W-D and PHOEBE), our model offers substantially improved computational efficiency. Comparing with previous studies (e.g., \citealt{2022AJ....164..200D, 2025ApJS..277...51L}), our model demonstrates superior capability in simultaneously solving multiple-band LCs and determining all four spot parameters. By applying our model to both PHOEBE-generated LCs and eight well-studied systems from the literature, we demonstrate that it yields robust solutions. We then analyzed the V- and I-band LCs of OGLE contact binaries using our NN model, deriving physical parameters for 3,541 systems.

To facilitate researchers to utilize our code, we have packaged it into a user-friendly Windows executable (CBLA.exe) file for convenient deployment. The executable file has been uploaded to and is available for download from the China-VO data repository (\url{https://doi.org/10.12149/101626}). Our program can be applied to quickly analyze the LCs of contact binaries discovered by almost all current photometric sky surveys, as well as most telescope observations. This capability is of great significance for investigating the formation, structure, and evolution of contact binary systems.

\begin{acknowledgments}
We thank the anonymous reviewer for insightful comments and constructive suggestions, which have significantly improved the quality of this manuscript. This work is supported by National Natural Science Foundation of China (NSFC) (No. 12273018), and the Joint Research Fund in Astronomy (No. U1931103) under cooperative agreement between NSFC and Chinese Academy of Sciences (CAS), and by Taishan Scholars Young Expert Program of Shandong Province, and by the Qilu Young Researcher Project of Shandong University, and by the Young Data Scientist Project of the National Astronomical Data Center, and by the Cultivation Project for LAMOST Scientific Payoff and Research Achievement of CAMS-CAS. The calculations in this work were carried out at Supercomputing Center of Shandong University, Weihai.
\end{acknowledgments}


\bibliography{sample7}{}
\bibliographystyle{aasjournalv7}

\end{document}